\documentclass{elsart}
\usepackage{epsf,amssymb}
\journal{Physics Letters A}
\begin{document}
\newcommand{\bra}{\langle}
\newcommand{\ket}{\rangle}
\newcommand{\eq}[2]{\begin{equation}\label{#1} #2 \end{equation}}
\newcommand{\tbf}[1]{{\bf #1}}
\newcommand{\tit}[1]{{\it #1}}
\newcommand{\intRR}{\int\limits_{-\infty}^{\infty}}
\newcommand{\intR}{\int\limits_{0}^{\infty}}
\newcommand{\intRt}{\int_{0}^{\infty}}
\newcommand{\LRD}[1]{\frac{{{\displaystyle\leftrightarrow}
\atop {\displaystyle\partial}}}{\partial #1}}
\newcommand{\lrd}[1]{\stackrel{\displaystyle \leftrightarrow}
{\displaystyle\partial_{#1}}}
\newcommand{\lrx}[1]{\stackrel{\displaystyle \leftrightarrow}
{\displaystyle #1}}
\newcommand{\lrX}[1]{\stackrel{\displaystyle \longleftrightarrow}
{\displaystyle #1}}
\newcommand{\diff}[1]{\partial /{\partial #1}}
\newcommand{\diffm}[2]{\partial#1/\partial #2}
\newcommand{\Diff}[1]{\frac{\partial}{\partial #1}}
\newcommand{\hc}[1]{#1^{\dagger}}
\newcommand{\DiffM}[2]{\frac{\partial #1}{\partial #2}}
\newcommand{\SDiffM}[2]{\frac{\partial^2 #1}{\partial {#2}^2}}
\newcommand{\ds}{\displaystyle}
\newcommand{\sign}{\,{\rm sign}\,}
\textheight=23cm
\begin{frontmatter}
\title{Pair creation by homogeneous electric field from the
point of view of an accelerated observer}
\author{N.B. Narozhny}\ead{narozhny@theor.mephi.ru},
\author{V.D. Mur},
\author{A.M. Fedotov\corauthref{cor}},
\corauth[cor]{Corresponding author} \ead{fedotov@cea.ru}
\address{Moscow Engineering Physics Institute (state university), 115409 Moscow, Russia}
\date{}

\begin{abstract}
With a special choice of gauge the operator of the
Klein-Fock-Gordon equation in homogeneous electric field respects
boost symmetry. Using this symmetry we obtain solutions for the
scalar massive field equation in such a background (boost modes in
the electric field). We calculate the spectrum of particles
created by the electric field, as seen by an accelerated observer
at spatial infinity of the right wedge of Minkowski spacetime. It
is shown that the spectrum and the total number of created pairs
measured by a remote uniformly accelerated observer in Minkowski
spacetime are precisely the same as for inertial observers.

\end{abstract}

\end{frontmatter}

\textbf{1.} The Schwinger process of pair creation by a constant
electric field from vacuum \cite{Schwinger} have been studied in
details (using different methods and different gauges of the
external field) in the early 70-s \cite{N,NN,VSP,GMF,TMF,PLI}. In
this literature the quantum states of created particles were
labelled either by values of generalized momentum (non-stationary
gauge) or energy (stationary gauge). However the constant electric
field permits another symmetry which is based on invariance of the
field with respect to Lorentz transformations along its direction.
If the field is directed along the $z$-axis, this symmetry is
generated by the boost operator
$\mathcal{B}=i(t\diff{z}+z\diff{t})$. The presence of boost
symmetry in the problem makes it very similar to the problem of
particle creation by an eternal black hole \cite{Unruh,BD}. Thus
treatment of the Schwinger process in terms of boost modes is of
great importance for the quantum field theory in curved spacetime.

Besides providing a possible algorithm for investigation of the
process of particle creation in the background of eternal black
hole, the boost modes based approach to the Schwinger problem is
of independent interest. It is very natural to use Milne--Rindler
coordinate map (see, e.g., \cite{BD}) in problems with boost
symmetry. Physically transition to these coordinates means
transition to a non-inertial reference frame, namely, to a
uniformly accelerated reference frame in the right and left wedges
of Minkowski spacetime (MS). We know the average number of
particles created by a constant electric field as it is seen in
inertial reference frames. Therefore calculation of this quantity
measured by an accelerated observer will clear up the question of
how inertial forces influence the spectrum and the total number of
created particles.

\textbf{2.} We consider the Schwinger problem for massive scalar
particles where for the electric field $E$ we chose a special
gauge \eq{gauge}{A_{\mu}=-(E/2)\epsilon_{\mu\nu}x^{\nu}, \quad
x^{\mu}=\{t,z\}, \,\; \epsilon_{\mu\nu}=-\epsilon_{\nu\mu}, \;
\epsilon_{tz}=1,} which provides invariance of the
Klein-Fock-Gordon (KFG) equation under boost
transformations\footnote{In the current paper we use natural units
$\hbar=c=1$ and metric with the signature $(+,-)$. Note that the
proposed gauge is very similar to the well-known gauge
$\mathbf{A}=(1/2)\mathbf{H}\times\mathbf{r}$, which provides
explicit rotational symmetry of the Schr\"odinger equation in the
presence of a uniform magnetic field.}. The latter equation in the
gauge (\ref{gauge}) takes the form \eq{KFG}{
\left(\square-eE\mathcal{B}+\frac14
e^2E^2x_+x_-+m^2\right)\phi(x)=0,} where $e$ and $m$ are the
electric charge and mass of the particle, $x_{\pm}=t \pm
z,\;\square = 4\partial^2/\partial x_+
\partial x_-$ and the boost generator
$\mathcal{B}=\mathcal{D}_{(+)} -\mathcal{D}_{(-)}, \;
\mathcal{D}_{(\pm)} =ix_{\pm}
\partial/\partial x_{\pm}.$ One can easily check that the
KFG operator at the l.h.s. of Eq.(\ref{KFG}) commutes with
$\mathcal{B}$. Therefore solutions for equation (\ref{KFG}) can be
labelled by eigenvalues $\kappa$ of the boost generator
$\mathcal{B}$. We will call the solutions to Eq.(\ref{KFG})
$\phi_{\kappa}(x)$ satisfying the equation
$\mathcal{B}\phi_{\kappa}=\kappa\phi_{\kappa}$ "the boost modes in
the electric field".

\begin{figure}[hhttbb]
\epsfxsize=13cm \centerline{\epsffile{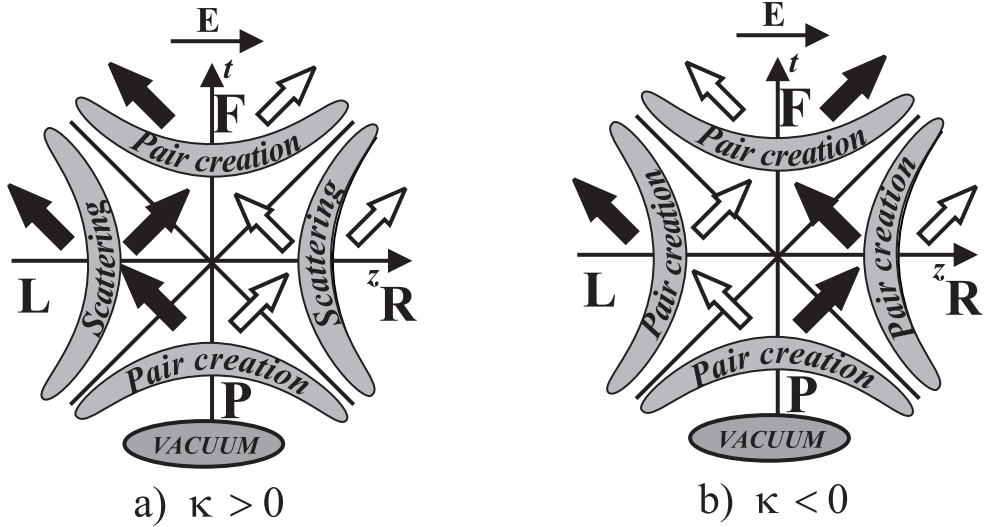}}
\caption{Spacetime location of particle creation regions and
direction of particle (empty arrows), antiparticle (filled arrows)
fluxes for different signs of $\kappa$.} \label{scheme}
\end{figure}

The commuting boost $\mathcal{B}$ and dilatation
$\mathcal{D}=\mathcal{D}_{(+)} +\mathcal{D}_{(-)}$ vector fields
generate local pseudo\-eucli\-dean coordinates which admit
separation of variables in Eq. (\ref{KFG}). In the future ($F$)
and past ($P$) wedges of MS, see Fig.\ref{scheme}, these are Milne
coordinates \eq{Mln}{\tau=\pm m(x_+x_-)^{1/2}, \;
\sigma=1/2\ln(x_+/x_-),} while in the Rindler right ($R$) and left
($L$) wedges, Rindler coordinates
\eq{Rnd}{\eta=1/2\ln(-x_+/x_-),\;\rho=\pm m(-x_+x_-)^{1/2},}
(compare to \cite{BD}). The four regions $R$, $L$, $P$ and $F$ are
bounded by horizons which belong to the light cone $x^2\equiv
t^2-z^2=0$, and (together with the horizons and the origin $x=0$)
cover the whole MS. Since the coordinate lines $\rho = const$ in
the Rindler wedges $R$ and $L$ coincide with the world lines of
uniformly accelerated observers in MS the resulting reference
frame can be considered as a uniformly accelerated one.

\textbf{3.} It was shown in Refs. \cite{NN,FIAN} that classical
solutions for the relevant field equation contain complete
information on the pair creation process. Here we will follow
these papers in our analysis instead of constructing a consistent
second quantized theory. We will assume that the initial state of
the field in $P$-wedge at far past ($\tau \rightarrow -\infty,$ or
$x_{\pm}\rightarrow -\infty$) is a vacuum state.

It follows from \cite{NN,FIAN} that there exist two non-equivalent
complete sets of modes in the Schwinger problem. It was shown in
\cite{FIAN} that complete information on the pair creation process
can be derived from any solution belonging to any of the sets. It
is convenient for us to use the solution which in the $P$-wedge
takes the form \eq{bm}{
\begin{array}{c}\ds
\phi_{\kappa}(x)=\frac{1}{\sqrt{2}}(-mx_-)^{i\kappa}
\exp\left\{-\frac{i}{4}\,\mathcal{E}m^2x_+
x_--\frac{\pi}{4\,\mathcal{E}} \right\}\times
\\ \\ \ds
\times
\Psi\left(\frac12+\frac{i}{2\,\mathcal{E}}\,
,\,1+i\kappa\,,\frac{i}{2}\, \mathcal{E}m^2x_+x_-\right),
\end{array}}
where $\mathcal{E}=E/E_{cr}$ is the electric field strength in the
units of critical QED field $E_{cr}=m^2c^3/e\hbar$, and
$\Psi(a,c,\zeta)$ is the Tricomi function \cite{BE1}. The explicit
form for the solution (\ref{bm}) in other wedges may be obtained
by analytic continuation in such a way that for transitions from
one wedge to another one should use substitutions $-x_{\pm}
\rightarrow x_{\pm}e^{-i\pi},$ (compare to Ref. \cite{PRD}). At
the limit $E \rightarrow 0$ the modes (\ref{bm}) coincide up to a
phase factor with the boost modes $\;\Psi^*_{-\kappa}(x)$ for
empty MS, see Ref. \cite{PRD}.

Let us first analyze the particle creation process in the
$P$-wedge. Using the asymptotic expression for the Tricomi
function \cite{BE1} we can see that at far past
($x_{\pm}\to-\infty$) the solution (\ref{bm}) acquires
semiclassical form \eq{past}{
\phi_{\kappa}(x)=(2\,|\mathcal{D}S_{\kappa}|)^{-1/2}
\exp(iS_{\kappa}),} where
$$
\ds S_{\kappa}(x_+,x_-)=-\frac{\mathcal{E}}{4}\,m^2x_+x_-
-\frac{1}{2\,\mathcal{E}}\ln\left(m^2x_+x_-\right)+\kappa\ln(-mx_-)
+\ldots
$$
is the classical action of a particle with a charge $e$ moving in
the electric field (\ref{gauge}). This solution is normalized by
the condition $\mathcal{J}^{\,\tau}=-e$ which means that the
charge per unit interval of $\sigma$ in the local reference frame
is equal to $-e$. Hence, in the in-region the solution (\ref{bm})
describes an incoming antiparticle accelerated by the electric
field $E$. Here
$$
\mathcal{J}^{\mu}=e\sqrt{-g}g^{\mu\nu}
\left(i\phi_{\kappa}^*\LRD{x^{\nu}}\varphi_{\kappa}-
2eA_{\nu}\phi_{\kappa}^*\phi_{\kappa}\right),
$$
is the vector density of current, $\tau$-component of which for
the wedge $P$ can be represented in the form
$\;\mathcal{J}^{\,\tau}=-e\phi_{\kappa}^*\!\!\lrx{\mathcal{D}}\!\!
\phi_{\kappa}$.

Near the light cone $x_+x_-=0$ the solution (\ref{bm}) reduces to
\cite{BE1} \eq{lc}{
\begin{array}{r}\ds
\phi_{\kappa}(x)\sim \frac1{\sqrt{2}}e^{-\pi /4\,\mathcal{E}}
\frac{\Gamma(-i\kappa)}{\Gamma(1/2+i/2\,\mathcal{E}-i\kappa)}(-mx_-)^{i\kappa}
+\\ \\ \ds + \frac1{\sqrt{2}}e^{\pi\kappa/2-\pi/4\,\mathcal{E}}
\frac{\Gamma(i\kappa)}{\Gamma(1/2+i/2\,\mathcal{E})}
\left(-\frac{\mathcal{E}}{2}mx_+\right)^{-i\kappa}.
\end{array}}

The first term at the r.h.s. is obviously a wave propagating to
the right, while the second term is a wave propagating to the
left, both waves travelling with the speed of light. Near the
horizon the charge densities carried by the right- and left-going
waves are respectively given by \eq{j+}{\begin{array}{cc}\ds
\mathcal{J}^{\,\tau}_{(-)}=-ie\phi_{\kappa}^*\lrx{D}_{(-)}\phi_{\kappa}=
\,e\frac{1+e^{2\pi\kappa-\pi/\mathcal{E}}}{e^{2\pi\kappa}-1},\label{j-}\\
\\ \ds \mathcal{J}^{\,\tau}_{(+)}
=-ie\phi_{\kappa}^*\lrx{D}_{(+)}\phi_{\kappa}=
-e\frac{1+e^{-\pi/\mathcal{E}}}{1-e^{-2\pi\kappa}}.
\end{array}}

It is easily seen that $\mathcal{J}^{\,\tau}_{(-)}>0$ and
$\mathcal{J}^{\,\tau}_{(+)}<0$ if $\kappa>0$. This means that the
right-going wave describes the flux of particles created by the
electric field, while the left-going wave describes the flux of
created antiparticles together with the incoming one. Thus, these
charge densities should be written in the form, compare
\cite{NN,FIAN} $\mathcal{J}^{\,\tau}_{(-)}=en_{\kappa}^{(P)}$,
$\mathcal{J}^{\,\tau}_{(+)}=-e\left(1+n_{\kappa}^{(P)}\right)$,
where $n_{\kappa}^{(P)}$ is the number of pairs in the $\kappa$-th
mode, created by the electric field from vacuum in the $P$-wedge.
Obviously, we have: \eq{kappa>0}{
n_{\kappa}^{(P)}=\frac{1+e^{2\pi\kappa-\pi/\mathcal{E}}}{e^{2\pi\kappa}-1},
\quad\kappa>0.} If $\kappa<0$, $\mathcal{J}^{\,\tau}_{(-)}<0$ and
$\mathcal{J}^{\,\tau}_{(+)}>0$. Therefore in this case the
right-going wave describes the flux of antiparticles, while the
left-going wave describes the flux of particles, and the
corresponding charge densities should be written in the form
$\mathcal{J}^{\,\tau}_{(-)}=-e\left(1+n_{\kappa}^{(P)}\right)$,\,
$\mathcal{J}^{\,\tau}_{(+)}=e n_{\kappa}^{(P)}$ with \eq{kappa<0}{
n_{\kappa}^{(P)}=\frac{1+e^{-\pi/\mathcal{E}}}{e^{2\pi|\kappa|}-1},
\quad\kappa<0.} Thus we conclude that for $\kappa>0$ particles
created in the $P$-wedge travel to the $R$-wedge and the
antiparticles to the $L$-wedge, while for negative values of
$\kappa$ we have the opposite situation.

\textbf{4.} Now let us consider what happens in the $R$-wedge. In
terms of Rindler coordinates (\ref{Rnd}) the KFG equation
(\ref{KFG}) after separation of the time-like variable $\eta,\;
\phi_{\kappa}=e^{-i\kappa\eta}\varphi_{\kappa}(\rho),$ formally
coincides with the stationary Schr\"odinger equation with respect
to the independent variable $u=\ln{\rho}$  \eq{RW_eq}{
\left[-\frac{d^2}{du^2}+
U_{\kappa}\right]\varphi_{\kappa}=\kappa^2\varphi_{\kappa},} where
the effective potential reads \eq{eff_pot}{
U_{\kappa}(\rho)=(1-\mathcal{E}\kappa)\rho^2-\frac14\,\mathcal{E}^2\rho^4,
\quad \rho=e^u.} If $\kappa\ge 1/\mathcal{E}$, the effective
potential is monotonously decreasing with $\rho^2$. If
$\kappa<1/\mathcal{E}$, then $U_{\kappa}(\rho)$ is a potential of
barrier type with maximum $U_{\kappa}^{(m)}$ at $\rho=\rho_m$
\eq{max}{ \rho_m^2=2\mathcal{E}^{-1}(\mathcal{E}^{-1}-\kappa),
\quad U_{\kappa}^{(m)}=(\mathcal{E}^{-1}-\kappa)^2.} However real
solutions for the equation $U_{\kappa}(\rho)=\kappa^2$ exist only
for $\kappa<1/2\,\mathcal{E}$. It means that the values
$\kappa>1/2\,\mathcal{E}$ correspond to above-barrier scattering,
while for $\kappa<1/2\,\mathcal{E}$ the Eq.(\ref{RW_eq}) describes
the sub-barrier tunnelling. The classical turning points of the
potential for the sub-barrier situation are located at
\eq{tp}{\rho_{\pm}=\left\{\begin{array}{ll}\mathcal{E}^{-1}
(1\pm\sqrt{1-2\kappa\, \mathcal{E}}\,),& \quad
0<\kappa<1/2\,\mathcal{E}\,, \\
\mathcal{E}^{-1}(\sqrt{1+2|\kappa|\,\mathcal{E}}\pm 1), &\quad
\kappa<0.\end{array}\right.}We will again base our analysis on the
solution (\ref{bm}) though analytically continued to the
$R$-wedge. It consists of a superposition of in- and outgoing
waves near horizons ($u\rightarrow -\infty$) while at the right
spatial infinity of the wedge ($u\rightarrow \infty$) reduces to a
semiclassical right-going wave. The transmission $D_{\kappa}$ and
reflection $R_{\kappa}$ coefficients can be obtained by the
standard quantum mechanical procedure and are equal to \eq{DR}{
D_{\kappa}=\frac{1-e^{-2\pi\kappa}}{1+e^{\pi/\mathcal{E}-2\pi\kappa}}\,,\quad
R_{\kappa}=1-D_{\kappa}=
\frac{1+e^{-\pi/\mathcal{E}}}{1+e^{2\pi\kappa-\pi/\mathcal{E}}}\,.}
The scattering interpretation of the coefficients (\ref{DR}) is
consistent for all positive values of $\kappa$. However, it
follows from Eq.(\ref{DR}) that $D_{\kappa}<0$ and $R_{\kappa}>1$
if $\kappa<0$. This phenomena is known as the Klein paradox
\cite{Klein} (see also \cite{NN,FIAN}) and is explained by the
effect of pair creation by the external field. The fact that pairs
are created only with $\kappa<0$ is easy to understand if one
notes that, as it follows from Eqs.(\ref{tp}), the work of the
external field at the sub-barrier region is equal to
$$A=m\,\mathcal{E}(\rho_+-\rho_-)=\left\{
\begin{array}{ll}
2m\sqrt{1-2\kappa\,\mathcal{E}},&
\quad0<1/\kappa<1/2\,\mathcal{E},
\\  2m, & \quad\kappa<0,
\end{array}
\right.$$ and is sufficient for pair creation only for $\kappa<0.$
According to Ref.\cite{FIAN} (see also \cite{Hund}) the number of
created pairs in this situation is given by the absolute value of
the transmission coefficient \eq{n_R}{
n_{\kappa}^{(R)}=\theta(-\kappa)|D_{\kappa}|=\theta(-\kappa)
\frac{e^{2\pi|\kappa|}-1}{e^{\pi/\mathcal{E}+2\pi|\kappa|}+1}.} It
is important that this result is valid in the $R$-wedge only under
assumption that the external field creates pairs from {\it
vacuum}. In the second quantized theory it means that there exists
the amplitude of vacuum -- vacuum transition $_r\bra 0|0\ket_l$,
where $|0\ket_l$ is the state of the field without right-going
particles on the left of the barrier and $|0\ket_r$ is the state
without left-going particles on the right of it. It is clear that
in our case the state $|0\ket_l$ could be prepared only if no
particles arrive to the $R$-wedge from outside. In other words,
the validity of the spectrum (\ref{n_R}) requires zero boundary
condition for the field in $R$-wedge at $u\rightarrow -\infty$, or
$\rho\rightarrow 0$ (compare with \cite{PRD}). However there is no
reason for implementation of such boundary condition in MS. We
will consider now what is happening in the $R$-wedge with regard
for the fact that the $R$-wedge is a part of MS but not a separate
spacetime.

\textbf{5.} The total picture can be reconstructed from what we
already know about $P$- and $R$-wedges. Physical phenomena in the
$F$- ($L$-) wedge are identical to those in $P$- ($R$-) wedge due
to the symmetry of the problem with respect to time reversal (or
space inversion). This total picture is shown in
Fig.~\ref{scheme}.
In $P$- and $F$-wedges the external field creates pairs
characterized by $\kappa$ of both signs. The particles created in
the $F$ wedge stay within it, while the particles created in the
$P$-wedge go to the $R$- and $L$-wedges. If $\kappa>0$, the
particles get to the $R$-wedge, while the antiparticles to the
$L$-wedge. The incoming particles (antiparticles) are partially
reflected there by the effective potential (\ref{eff_pot}) and
arrive to the $F$-wedge. Those particles (antiparticles) which
penetrate through the potential go to the right infinity of the
$R$-wedge (left infinity of the $L$-wedge), where they can be
detected by a remote uniformly accelerated (Rindler) observer who
is moving along a hyperbolic world line $\rho={\rm const}$. In
particular, the Rindler observer in the right wedge will detect
created particles and will not observe antiparticles. Since the
particles behind the scattering barrier in the $R$-wedge are
continuously accelerated by the electric field, they are
travelling with almost the speed of light. It is clear that the
number of particles with $\kappa>0$ which are crossing the world
line of a remote Rindler observer per unit his local time is given
by \eq{J_p}{
J_{\kappa}=D_{\kappa}n_{\kappa}^{(P)}=e^{-\pi/\mathcal{E}},\quad
\kappa>0.}

If $\kappa<0$, then the particles created in the wedge $P$ get to
the $L$-wedge, while the antiparticles to the $R$-wedge. All of
them are reflected by the effective potential (\ref{eff_pot}) and
follow to the $F$-wedge. But the electric field creates pairs with
$\kappa<0$ directly in the $R$ and $L$ wedges. The particles
created in the wedge $R$ (antiparticles in $L$) move to the right
(left) spatial infinity, while the antiparticles (particles) go to
the $F$-wedge. Hence, independently of the sign of $\kappa$ the
right Rindler observer always detects only particles. But if at
$\kappa>0$ these particles originate from the $P$-wedge, at
$\kappa<0$ they are created at the sub-barrier region in the
$R$-wedge itself. The important point is that the number of
particles created in the wedge $R$ (and $L$) in the current
situation differs from the expression (\ref{n_R}) because the
latter is related to particle production from vacuum. However,
pairs in the wedge $R$ are created in the presence of
antiparticles arrived from the $P$-wedge. Since scalar particles
satisfy the Bose statistics, presence of antiparticles results in
stimulation of pair production in the $R$-wedge. Therefore the
number of particles with $\kappa<0$ actually produced in the wedge
$R$ and equal to the number of particles detected by the remote
Rindler observer per unit his local time is given by (see, e.g.,
Ref.~\cite{FIAN}) \eq{J_n}{
J_{\kappa}=n_{\kappa}^{(R)}\left(1+n_{\kappa}^{(P)}\right)=
e^{-\pi/\mathcal{E}},\quad \kappa<0,} with the quantity
$n_{\kappa}^{(R)}$ defined in Eq.(\ref{n_R}). Thus, the spectrum
of particles measured by the Rindler observer is independent of
quantum number $\kappa$.

It is worth noting that this result can also be obtained directly
from the solution (\ref{bm}). Indeed, the asymptotic form
(\ref{past}) of this solution can be analytically  continued to
the right spatial infinity of the $R$-wedge along the path in the
complex space $\{x_{+},x_{-}\}$, which goes round the horizon
$x_{+}=0$ at sufficiently large distance, such that the
semiclassical approximation remains valid along all the path. It
is easy to see, that after such continuation the classical action
$S_{\kappa}(x_{+},x_{-})$ acquires constant imaginary part $\Delta
S_{\kappa}=i\pi/2\,\mathcal{E}$. Since the solution (\ref{bm})
corresponds to an incoming antiparticle at far past in wedge $P$
and anti\-particles do not penetrate through the potential
$U_{\kappa}$, see Fig. \ref{scheme}, the asymptotic of solution
(\ref{bm}) describes at the right spatial infinity of the
$R$-wedge only particles created by the electric field. Hence, the
asymptotic of the current in this region is equal to $eJ_{\kappa}$
and after simple calculations we arrive to the result
$J_{\kappa}=\exp\{-2\,{\rm Im}\,S_{\kappa}\}$ in full agreement
with Eqs. (\ref{J_p}), (\ref{J_n}).

Our analysis can be generalized to the 3-dimensional case. Indeed,
it is clear that the 3-dimensional KFG equation after separation
of variables orthogonal to the direction of the electric field
reduces to the equation for the 1-dimensional problem with the
effective mass $\sqrt{m^2+p_{\perp}^2}$ instead of $m$. Hence the
number of particles with quantum numbers
$(\kappa,\mathbf{p}_{\perp})$ detected by remote Rindler observer
per unit local time $\eta$ can be obtained from Eqs. (\ref{J_p}),
(\ref{J_n}) by substitution $m \rightarrow \sqrt{m^2+p_{\perp}^2}$
\eq{3d}{ J_{\kappa,\mathbf{p}_{\perp}}=
\exp\left(-\frac{\pi\left(m^2+p_{\perp}^2\right)}{eE}\right).}

\textbf{6.} Finally, we summarize our results and present the
conclusions. In this paper we have analyzed the process of pair
creation by a constant homogeneous electric field in MS viewed
from a uniformly accelerated reference frame. We have shown that
the spectrum of particles (antiparticles) measured by a remote
uniformly accelerated observer is precisely the same as the one
measured by a conventional inertial observer (see, e.g., Refs.
\cite{NN,FIAN}). This means that inertial forces cannot create
particles. In particular, no particles can be observed at spatial
infinity in the absence of the electric field with respect to both
reference frames. In our opinion, this conclusion is an additional
explicit demonstration of non-existence of the so-called Unruh
effect \cite{Unruh}. Previously, we have come to the same
conclusion in Refs. \cite{PRD,Ann,det} on the basis of quite
different arguments.

Let us emphasize that the number of pairs (\ref{n_R}) created in
the $R$-wedge, when it is considered as a separate spacetime with
a special condition imposed at its boundary, is absolutely
different from the one (\ref{J_p}), (\ref{J_n}) measured by a
remote uniformly accelerating observer in MS where communication
between the wedges is possible. This result provides additional
evidence for the necessity of boundary conditions for consistent
field quantization on incomplete manifolds, see Refs.
\cite{Ann,PRD}.

The authors wish to thank L.B. Okun for constant interest to our
work. We appreciate support from the Russian Fund for Basic
Research and Ministry of Education of Russian Federation.

\end{document}